\documentclass[aps,showpacs,amsmath,amssymb]{revtex4}

\usepackage[dvips]{graphicx}
\usepackage{bm}
\usepackage{hyperref}

\newcommand{\bra}[1]{\mbox{$\left\langle{#1}\right|$}}
\newcommand{\ket}[1]{\mbox{$\left|{#1}\right\rangle$}}

\begin{document}

\title{Hybrid protocol of remote implementations of quantum operations}
\author{Ning Bo Zhao, An Min Wang\footnote{Email Address: anmwang@ustc.edu.cn}}
\affiliation{Quantum Theory Group, Department of Modern Physics\\
University of Science and Technology of China, Hefei 230026, People
Republic of China}

\begin{abstract}
We propose a protocol of remote implementations of quantum operations by hybridizing bidirectional quantum state teleportation's (BQST) \cite{hpv01} and Wang's one \cite{wang06}. 
The protocol is available for remote implemetations of quantum operations in the restricted sets specified in Sec. \ref{new}.
We also give the proof of the protocol and point out its optimization.
As an extension, this hybrid protocol can be reduced to BQST and Wang protocols.
\end{abstract}


\maketitle
\section{Introduction}
Entanglement has been used as a key resource in many tasks in quantum information processing \cite{Plenio98}.
As a famous example of these tasks, quantum-state teleportation \cite{six93} means that an unkown quantum state is transferred among distant parties without physically sending the particle. 
Another important task is teleportation of a quantum operation \cite{hpv01}, where instead of an unkown state, an unknown quantum operation is transferred without physically sending the device.
If the teleported operation acts also on a remote unknown state, this task can also be called ``remote implementation of operation'' .
Recently, researches on this aspect are made on both theory \cite{hpv01}\cite{hpv02}\cite{wang06} and experiment \cite{guo04}\cite{guo05}\cite{hpguo05}.

When the operation is completely unknown, this remote implementation has to be completed via so-called bidirectional quantum state teleportation (BQST) \cite{hpv01}, in which the receiver teleports his target state to the sender, then after applying the operation, the sender teleports it back to the receiver. 
Apparently,  only a pair of quantum-state teleportations and one local quantum operation are implemented, and the required entanglement resources double that in quantum-state teleportations.

It is very interesting when the operation is partially unknown.
Here, ``partially unknown'' quantum operations mean they are belonging to some restricted sets that satisfy some given restricted conditions.
There are some protocols via which the partially unkown operation can be remote implemented using less resources than via BQST.
In other words, any operation in the restricted set with respect to a protocol can be remote implemented via this protocol.
Entanglement is a scarce resource in quantum information processing, and is more expensive than classical resources such as classical communication.
So, these protocols need also use entanglement resources as little as possible, and this economization is not insignificant.

In the case of one qubit operations, there are two such restricted sets, and operations in either of them can be teleported via a protocol (HPV) using the least entanglement resources \cite{hpv02}. 
These two restricted sets are one set which consists of digonal operations and one which consists of antidigonal operations.
Then in fact, HPV protocol may be considered as a group of two sub protocols with these two restricted sets respectively.
In HPV protocol, only one \emph{e}-bit of entanglment resources are required, and this is optimal.

These results have been developered for multiqubits cases\cite{wang06}.
Operations in anyone of the restricted sets in which there is just one none zero element in any row and any column, can be teleported via an extended protocol (Wang) using the least entanglement resources.
In the case of \emph{N}-qubit oprations, there are $2^N!$ such restricted sets, and $N$ \emph{e}-bits are required in Wang protocol, and this is optimal too.
Furthermore, HPV protocol is apparently a special case of Wang's when $N=1$.

The restricted sets in Wang protocol are matrices that has just one none zero element in any column or any row.
If the none zero elements are replaced by full rank squre matrices which have the same order each other, can we find protocols via which the operations can be teleported using the least entanglement resources?
In this paper, we will propose a protocol by hybriding Wang protocol and BQST, and furthermore, as their generalization and combination, it can be reduced to Wang protocol, HPV protocol and BQST.
This protocol will work when the restricted sets are $2^N \times 2^N$ block matrices which has just one none zero block in any column or any row, and every block of which is a $2^M \times 2^M$ full rank matrix.

This paper is organized as follows. 
We will introduce HPV protocol and Wang protocol firstly in Sec. \ref{hpv}.
Then, we will specify our new protocol and point out its optimality in Sec. \ref{new}. 
We summarize our conclusions and discuss some problems in Sec. \ref{con}.
In Apprendix \ref{prove}, we will give the proof of the new protocol.

\section{HPV protocol and Wang protocol}\label{hpv}
In the scene of remote implementation of quantum operations, Alice is set as a sender and Bob is set as a receiver.
Alice has the device that implement the local operation, and Bob has the unknown qubits to be operated.
They also share the necessary entanglement resources and have some accessorial qubits that assist them to accomplish the object.
As a result of an available protocol, Bob must finally get the qubits whose state is the same as Bob's initial qubits' state operated directly by the operation under the precondition without noise channel.
Furthermore, the protocol must involve only local quantum operations and classical communication (LOCC).
\subsection{HPV Protocol}
In Ref. \cite{hpv02}, the authors proposed the remote implementation of a quantum operation in some given restricted set.
They studied the case of one-qubit operations, and propose a simple but available protocol (HPV), and demostrated its optimality.
In the simplified HPV protocol, the initial state of the joint system of Alice and Bob is
\begin{equation}
|\Psi_{ABY}^{\rm ini}\rangle=|\Phi^+\rangle_{AB} \otimes |\xi\rangle_{Y},
\end{equation}
where
\begin{equation}
|\Phi^+\rangle_{AB}=\frac{1}{\sqrt{2}}(|00\rangle_{AB}+|11\rangle_{AB}),
\end{equation}
is a Bell states that is shared by Alice and Bob.
The qubit at Alice's side named qubit \emph{A}, and the other at Bob's side named qubit \emph{B}.
The qubit \emph{Y}
\begin{equation}
|\xi\rangle_{Y}=y_0|0\rangle_Y+y_1|1\rangle_Y
\end{equation}
is the qubit to be operated at Bob's side, and it is entirely unkown, that is, it can be in any pure state.

The quantum operation to be remote implemented belongs to one of the following two restricted sets
\begin{equation}
U(0)=\left( \begin{array}{cc}
u_{00} & 0 \\
0 & u_{11} 
\end{array} \right),\qquad 
U(1)=\left( \begin{array}{cc}
0 & u_{01} \\
u_{10} & 0 
\end{array} \right).
\end{equation}
It means that HPV protocol works when the operation belongs to either of them.
We will use $U(d)$ $(d=0,1)$ to denote the opertaion to be remote implemented.
In every actual processing, $d$ can only be exactly one value, and it is kown by Alice.
Before the protocol starts, Alice should tell Bob the information of the restricted sets using one bit through classical communication.  

HPV protocol can be expressed as following steps.

\paragraph*{Step 1: Bob's preparation.} 
Bob first performs a controlled-NOT using qubit \emph{Y} as the control and qubit \emph{B} as the target. 
Then, he measures the qubit \emph{B} in the computational bases $\ket{b}_B\bra{b}$ $(b=0,1)$.
So, Bob's preparation operations can be written as 
\begin{equation}
\mathcal{P}_B(b)=\left(\ket{b}_B\bra{b}\otimes\sigma_0^Y\right)
\left(\sigma_0^B\otimes\ket{0}_Y\bra{0}
+\sigma_1^B\otimes\ket{1}_Y\bra{1}\right),
\end{equation}
where $\sigma_0$ is a $2\times 2$ identity matrix and $\sigma_i$ $(i=1,2,3)$ are the Pauli matrices.

\paragraph*{Step 2: Classical communication from Bob to Alice.} 
Bob tells Alice his measurement result $b$ using one classical bit via a classical communication channel.

\paragraph*{Step 3: Alice's sending.}
After receiving the classical bit $b$, Alice performs $\sigma_b$ on qubit \emph{A}, and then performs the operation $U(d)$ on qubit \emph{A}. 
Then, Alice performs a Hadamard transformation on qubit \emph{A}, and then measure it in the computational bases $\ket{a}_A\bra{a}$ $(a=0,1)$.
So, Alice's sending operations can be written as 
\begin{equation}
{\mathcal{S}}_A(a,b;d)=\left(\ket{a}_A\bra{a}\right)\left[H^A
U(d)\sigma_b^A\right],
\end{equation}
where 
\begin{equation}
H=\frac{1}{\sqrt{2}}\left(\begin{array}{cc}1 &1 \\ 1
& -1\end{array}\right).
\end{equation}
is the Hadamard transformation.

\paragraph*{Step 4: Classical communication from Alice to Bob.}
Alice tells Bob her measurement result $a$ using one classical bit via a classical communication channel.

\paragraph*{Step 5: Bob's recovery.}
After receiving Alice's bit $a$, Bob firstly performs $\sigma_d$ on qubit \emph{Y}, and then performs $r(a)$ on it.
Here, $r(a)=(1-a)\sigma_0+a\sigma_3$ is $\sigma_0$ when $a=0$, and is $\sigma_3$ when $a=1$.
So, Bob's recovery operations can be written as
\begin{equation}
{\mathcal{R}}_B(a;d)=\sigma_0^B\otimes (r(a)^Y \sigma_d^Y).
\end{equation}

It is easy to conclude that after all steps finished Bob's qubit \emph{Y} results in the state $U(d)\left(y_0\ket{0}_Y+y_1\ket{1}_Y\right)$.
This means that the protocol is faithful and determined.

All of the operations in the protocol can be jointly written as
\begin{equation}
{\mathcal{I}}_R(a,b;d)=
\left[\sigma_0^A\otimes{\mathcal{R}}_B(a;d)\right]
\left[{\mathcal{S}}_A(a,b;d)\otimes\sigma_0^B\otimes\sigma_0^Y\right]
\left[\sigma_0^A\otimes{\mathcal{P}}_B(b)\right].
\end{equation}
So, the processing of the protocol can be expressed as
\begin{equation}
\ket{\Psi_{ABY}^{\rm
final}(a,b;d)}={\mathcal{I}}_R(a,b;d)\ket{\Psi_{ABY}^{\rm ini}}=
\frac{1}{2}\ket{a b}_{AB}\otimes U(d)\ket{\xi}_Y.
\end{equation}

\begin{figure}[ht]
\begin{center}
\includegraphics[scale=0.60]{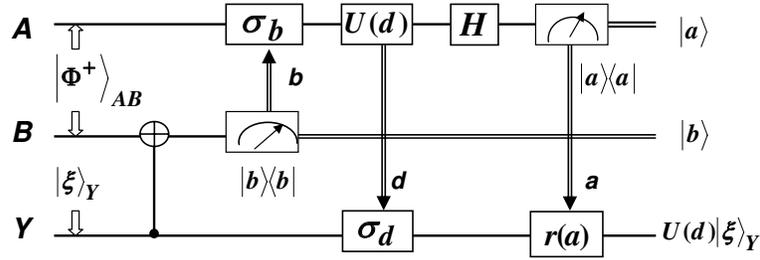}
\end{center}
\vskip -0.1in \caption{ \label{fig_hpv}
Quantum circuit of the HPV protocol, 
where $U(d)$ is the quantum operation to be remotely implemented, 
$H$ is the Hadamard gate, 
$\sigma_b,\sigma_d$ are identity matrices or {\sc not} gates ($\sigma_1$) with respect to $b,d=0$ or $b,d=1$, respectively, 
and $r(a)=(1-a)\sigma_0+a\sigma_3$ is an identity matrix when $a=0$ or a phase gate ($\sigma_3$) when $a=1$. 
``$\Rightarrow$'' indicates the transmission of classical communication.
} 
\end{figure}

We plot the quantum circuit of the HPV protocol in Fig. \ref{fig_hpv}.

\subsection{Wang Protocol}
Wang protocol deals with the case of multiqubits.
In the case of $N$ qubits, the initial state of Alice and Bob is
\begin{equation}
\ket{\Psi_N^{\rm
ini}}=\left(\bigotimes_{m=1}^N\ket{\Phi^+}_{A_mB_m}\right)
\otimes\ket{\xi}_{Y_1Y_2\cdots Y_N},
\end{equation}
where 
\begin{equation}
\ket{\xi}_{Y_1Y_2\cdots
Y_N}=\sum_{k_1,k_2,\cdots k_N=0}^1 y_{k_1k_2\cdots
k_N}\ket{k_1k_2\cdots k_N},
\end{equation}
is an arbitrary pure state.
Alice has the qubits $A_1A_2\cdots A_N$, and Bob has the qubits $B_1B_2\cdots B_N$ and $Y_1Y_2\cdots Y_N$.

The operation to be remote implemented is in one of the following $2^N!$ restricted sets 
\begin{equation}
T^r_N(x,t)=\sum_{m=1}^{2^N}t_m\ket{p_m(x),D}\bra{m,D},
\end{equation}
where $D$ indicates the decimal system, so $\ket{1,D}=\ket{00\cdots0}$, $\ket{2,D}=\ket{00\cdots1}$, $\ket{2^N,D}=\ket{11\cdots1}$, etc.
And, 
\begin{equation}
p(x)=\{p_1(x),p_2(x),\cdots,p_{2^N}(x)\},
\end{equation}
is a permutation of the list $\{1,2,\cdots,2^N\}$, where $x=1,2,\cdots,2^N!$ labels all of the $2^N!$ permutations, and also labels all of the restricted sets.
The initial state of qubits $Y_1Y_2\cdots Y_N$ can be similarly written as
\begin{equation}
\ket{\xi}_{Y_1Y_2\cdots Y_N}=\sum_{m=1}^{2^N} y_m\ket{m,D}.
\end{equation}
Similarly, Alice should tell Bob the information of the restricted set, so that Bob can choose the corresponding rescovery operation from Wang protocols.

Wang protocol can be expressed as following steps.

\paragraph*{Step 1: Bob's preparation.}
Bob first performs $N$ controlled-NOT respectively using qubits $Y_1Y_2\cdots Y_N$ as the controls and qubits $B_1B_2\cdots B_N$ as the targets. 
Then, he measures the qubits $B_1B_2\cdots B_N$ in the computational bases $\ket{b_m}_{B_m}\bra{b_m}$ $(b_m=0,1)$ $m=0,1,\cdots,N$.
So, Bob's preparation operations can be written as 
\begin{equation}
{\mathcal{P}}_B(b_1,b_2,\cdots,b_N)=
\bigotimes_{m=1}^N \left[(\ket{b_m}_{B_m}\bra{b_m})   C^{not}(Y_m,B_m)\right].
\end{equation}

\paragraph*{Step 2: Classical communication from Bob to Alice.} 
Bob transfers his mesurement results $b_1,b_2,\cdots,b_N$ to Alice using $N$ classical bits.

\paragraph*{Step 3: Alice's sending.}
After receiving the classical bits $b_1,b_2,\cdots,b_N$, Alice performs $\sigma_{b_m}$ on qubit $A_m$ respectively, and then performs the \emph{N}-qubit operation $T^r_N(x,t)$ on qubits $A_1A_2\cdots A_N$. 
Then, Alice performs a Hadamard transformation on qubit $A_m$ respectively, and then measure them respectively in the computational bases $\ket{a_m}_{A_m}\bra{a_m}$ $(a_m=0,1)$.
So, Alice's sending operations can be written as 
\begin{equation}
{\mathcal{S}}_A(a_1,b_1,a_2,b_2,\cdots,a_N,b_N;x,t)
=\left(\bigotimes_{m=1}^N \ket{a_m}_{A_m}\bra{a_m}\right)
  \left(\bigotimes_{m=1}^N H^{A_m}\right)
  T^r_N(x,t)^{A_1A_2\cdots A_N}   \left(\bigotimes_{m=1}^N\sigma_{b_m}^{A_m}\right).
\end{equation}

\paragraph*{Step 4: Classical communication from Alice to Bob.}
Alice tells Bob her measurement results $a_1,a_2,\cdots,a_N$ using N classical bits via classical communication channels.

\paragraph*{Step 5: Bob's recovery.}
Bob firstly performs ${R}_N(x)$ on qubits $Y_1Y_2\cdots Y_N$, and then performs $r(a_m)^{Y_m}$ respectively on them.
Here,
\begin{equation}
{R}_N(x)=T^r_N(x,0)=\sum_{m=1}^{2^N}\ket{p_m(x),D}\bra{m,D},
\end{equation}
is a \emph{N}-qubit transformation depended only on $x$.
That is, it depends only on the kind of restricted sets.
So, Bob's recovery operations can be written as
\begin{equation}
{\mathcal{R}}_B(a_1,a_2\cdots,a_N;x)
=\left(\bigotimes_{m=1}^Nr(a_m)^{Y_m}\right)  {R}_N(x)^{Y_1Y_2\cdots Y_N}.
\end{equation}

All of the operations in the protocol can be written as
\begin{eqnarray}
{\mathcal{I}}_R(a_1,b_1,a_2,b_2;\cdots,a_N,b_N;x,t)
&=&{\mathcal{R}}_B(a_1,a_2,\cdots a_N;x)\nonumber\\
& &\times {\mathcal{S}}_A(a_1,b_1,a_2,b_2,\cdots, a_N,b_N;x,t)\nonumber\\
& &\times {\mathcal{P}}_B(b_1,b_2,\cdots,b_N).
\end{eqnarray}
After the protocol is finished, the final state becomes
\begin{eqnarray}
 & &\ket{\Psi_N^{\rm
final}(a_1,b_1,a_2,b_2,\cdots,a_N,b_N;x)}\nonumber\\& &\quad =
{\mathcal{I}}_R(a_1,b_1,a_2,b_2;\cdots,a_N,b_N;x,t)\ket{\Psi_N^{\rm
ini}}
\\ & &\quad = \frac{1}{2^N}\left(\bigotimes_{i=1}^N\ket{a_i b_i}_{A_iB_i}\right)\otimes
T^r_N(x,t)\ket{\xi}_{Y_1Y_2\cdots Y_N}.
\end{eqnarray}
Thus, the protocol is faithful and determined, too.

It should be pointed out that these two protocols in this section are both available even if Bob's qubits that to be operated are in mixed state, because all operations in them are linear.
Thus, the qubits to be operated can be indeed general, whether they are in pure state or in mixed state. 
\section{Hybrid protocol in the case of $N+M$ qubits}\label{new}
Consider the following restricted sets of $N+M$ qubits operations
\begin{equation}
T_{N,M}^{r}(x,G)=\sum_{m=1}^{2^N} |p_m(x),D\rangle \langle m,D| \otimes G_m,
\end{equation}
where $G_m$s can be any $2^M \times 2^M$ full rank matrices.
They are similar to the restricted sets in Wang protocol, and just replace the numbers $t_m$s by the matrices $G_m$s.
So we can attempt to deal with the anterior operations of $N$-qubit similarly to Wang protocol, and deal with the posterior operations of $M$-qubit via BQST.
Because all of the operations in these protocols are linear, we can expect that this meathod be successful.
Then, this protocol could be called ``Hybrid Protocol'', and apparently, $N+2M$ Bell states are required. 
The protocol will be specified thereinafter in this section, and the full proof can be found in Appendix \ref{prove}.
Of course, in this prototol, Alice should firstly tell Bob the information of the restricted set, so that Bob can choose the corresponding rescovery operation just as in HPV protocol and in Wang protocol.

The initial state of Alice and Bob is
\begin{equation}
\ket{\Psi_{N,M}^{\rm ini}}
=\left(\bigotimes_{m=1}^{N+2M}\ket{\Phi^+}_{A_mB_m}\right)
\otimes\ket{\xi}_{Y_1Y_2\cdots Y_{N+M}},
\end{equation}
where $\ket{\xi}_{Y_1Y_2\cdots Y_{N+M}}$ is an arbitrary pure state.
Alice has the qubits $A_1A_2\cdots A_{N+2M}$, and Bob has the qubits $B_1B_2\cdots B_{N+2M}$ and $Y_1Y_2\cdots Y_{N+M}$.

The hybrid protocol can be expressed as following steps.

\paragraph*{Step 1: Bob's preparation.}
Bob's operations in this step is the same as in Wang protocol, that is:
\begin{equation}
{\mathcal{P}}_B(b_1,b_2,\cdots,b_N)=
\bigotimes_{m=1}^N \left[(\ket{b_m}_{B_m}\bra{b_m})   C^{not}(Y_m,B_m)\right],
\end{equation}

\paragraph*{Step 2: Classical communication and teleportations from Bob to Alice.}
In this step, Bob first tells Alice his measurement results, then teleports the qubits $Y_{N+1}Y_{N+2}\cdots Y_{N+M}$ to Alice's qubits $A_{N+1}A_{N+2}\cdots A_{N+M}$ respectively using the Bell states $A_{N+1}B_{N+1}A_{N+2}B_{N+2}\cdots A_{N+M}B_{N+M}$.

\paragraph*{Step 3: Alice's sending.}
In this step, Alice's operations is similar to Wang protocol.
She need only replace the operation $T^r_N(x,t)^{A_1A_2\cdots A_N}$ in Wang protocol by the operation $T^r_{N,M}(x,G)^{A_1A_2\cdots A_{N+M}}$.
Her operations can be expressed as 
\begin{equation}
{\mathcal{S}}_A(a_1,b_1,a_2,b_2,\cdots,a_N,b_N;x,G)
=\left(\bigotimes_{m=1}^N \ket{a_m}_{A_m}\bra{a_m}\right)
  \left(\bigotimes_{m=1}^N H^{A_m}\right)
  T^r_{N,M}(x,G)^{A_1A_2\cdots A_{N+M}}   \left(\bigotimes_{m=1}^N\sigma_{b_m}^{A_m}\right).
\end{equation}

\paragraph*{Step 4: Classical communication and teleportations from Alice to Bob.}
Alice first tells Bob her measurement results.
Then, she teleports the qubits $A_{N+1}A_{N+2}\cdots A_{N+M}$ to Bob's qubits $B_{N+M+1}B_{N+M+2}\cdots B_{N+2M}$ respectively using the Bell states $A_{N+M+1}B_{N+M+1}A_{N+M+2}B_{N+M+2}\cdots A_{N+2M}B_{N+2M}$.

\paragraph*{Step 5: Bob's recovery.}
Bob first does the same as in Wang protocol.
\begin{equation}
{\mathcal{R}}_B'(a_1,a_2\cdots,a_N;x)
=\left(\bigotimes_{m=1}^Nr(a_m)^{Y_m}\right)  {R}_N(x)^{Y_1Y_2\cdots Y_N}.
\end{equation}
Then, he performs $M$ additional swapping operations on the qubits $Y_{N+1}B_{N+M+1}Y_{N+2}B_{N+M+2}\cdots Y_{N+M}B_{N+2M}$ respectively.
The swapping operation can be expressed as
\begin{equation}
\mathcal{E} =\left(
\begin{array}{cccc}
1&0&0&0\\
0&0&1&0\\
0&1&0&0\\
0&0&0&1
\end{array}
\right),
\end{equation}
and apparently, $\mathcal{E}^{X,Y}$ just exchanges the states of qubits $X,Y$.
So, his operations in this step can be expressed as
\begin{equation}
{\mathcal{R}}_B(a_1,a_2\cdots,a_N;x)
=\left(\bigotimes_{n=1}^M \mathcal{E}^{Y_{N+n},B_{N+M+n}} \right)   \left(\bigotimes_{m=1}^Nr(a_m)^{Y_m}\right)  {R}_N(x)^{Y_1Y_2\cdots Y_N}.
\end{equation}

After the protocol is completed, the final state of qubits $Y_1Y_2\cdots Y_{N+M}$ becomes
\begin{equation}
\ket{\Psi_{N+M}^{\rm final}}_{Y_1Y_2\cdots Y_{N+M}}=T_{N,M}^{r}(x,G) \ket{\xi}_{Y_1Y_2\cdots Y_{N+M}}.
\end{equation}
To this end, the initial aim is accomplished, and the protocol is faithful and determined.

In this protocol, $N+2M$ \emph{e}-bits are required.
These entanglement resources are necessary for any protocol that can be used to faithfully teleport any operation in one of the restricted sets. 
This conclusion can be drawn using similar methods as in Ref. \cite{hpv02}.
We can also get it by considering the following set
\begin{equation}
S=T^r_N(x,t) \otimes V_M,
\end{equation}
where, $V_M$ can be any $M$-qubit operation.
Apparently, $S\subset T_{N,M}^{r}(x,G)$, so if a protocol is available for restricted set $T_{N,M}^{r}(x,G)$, it is also available for restricted set $S$.
But operations in $S$ are only direct products of an \emph{N}-qubit operation and an $M$-qubit operation.
In fact, remote implementations of such operations can be separated into two irrelevant parts, one is for the anterior $N$ qubits, the other is for the posterior $M$ qubits.
So, from the Ref. \cite{wang06} and \cite{hpv01}, any protocol that can be used to faithfully teleport any operation in set $S$ has to consume no less than $N+2M$ \emph{e}-bits entanglement resources.
Thus, our protocol is optimal in this case.
And, because our restricted sets $T_{N,M}^{r}(x,G)$s are not the forgoing trivial one, our protocol is nontrivial too. 

Furthermore, when $M=0$ our protocol deduces to Wang protocol, and when $N=0$ it deduces to BQST. 
Especially, when $M=0$ and $N=1$ it becomes HPV protocol.

\section{Conclusion and Discussion }\label{con}
In this paper, we consider the remote implementation of operations in the restricted sets that have a block form.
Operations in restricted sets like this can not be dealt with by any anterior protocol except for BQST.
But, too many entanglement resources are required if directly using BQST protocol.
We have proposed a protocol that can be used to deal with the case that the restricted sets have a form specified in anterior section.
Any anterior protocol can be regarded as a special case of this protocol.
Then we have pointed out that our protocol is optimal, that is, it consumes the least entanglment resources.

There are many other restricted sets that our protocol can not be used to deal with.
However, because all of the elementary quantum gates can be included in our restricted sets, after using Wang's combined protocol, this problem is not serious.
Perhaps in a process of remote implementation of a quantum algorithm, our protocols are enough.
Of course, further researches can be made on quantum operations structure to classify the restricted sets, and on new protocols for every class of restricted sets.
Our method would provide some clues on these researches.
Furthermore, our method could also be used to the combined and the controlled remote implementations in Ref. \cite{wang07}.

Remote implementations of quantum operations is a critical step for the implementation of quantum ditributing computation and teleportation-based models of quantum computation. 
Investigations on it can give helps to the researches of the forgoing issues.
\section*{Acknowledgments}

We acknowledge all the collaborators of our quantum theory group at the Institute for Theoretical Physics of our university. 
This work was funded by the National Natural Science Foundation of China under Grant No. 60573008.

\begin{appendix}
\section{proof of our protocol}\label{prove}
In this appendix, we prove the hybrid protocol proposed in Sec. \ref{new}, and some detailed technologies are similar to the Ref. \cite{wang06}

The initial state of the qubits $Y_1Y_2\cdots Y_{N+M}$ can always be expressed as
\begin{eqnarray}
& & \ket{\xi}_{Y_1Y_2\cdots Y_{N+M}} \nonumber \\
&=&\sum_{k_1,k_2,\cdots k_{N+M}=0}^1 z_{k_1k_2\cdots k_{N+M}}\ket{k_1k_2\cdots k_{N+M}}\nonumber \\
&=& \sum_{k_1,k_2,\cdots,k_{N}=0}^{1} y_{k_1,k_2,\cdots,k_{N}} |k_1,k_2,\cdots,k_{N}\rangle_{Y_1Y_2\cdots Y_{N}} \otimes |\eta_{k_1,k_2,\cdots,k_{N}}\rangle_{Y_{N+1}\cdots Y_{N+M}} \nonumber \\
&=&\sum_{m=1}^{2^N} y_{m} |m,D\rangle_{Y_1Y_2\cdots Y_{N}} \otimes |\eta_{m}\rangle_{Y_{N+1}\cdots Y_{N+M}},
\end{eqnarray}
where $|\eta_{k_1,k_2,\cdots,k_{N}}\rangle$s or $|\eta_{m}\rangle$s need not be orthogonal each other.
So, in the sense of swapping transformations, the initial state of the total system can be expressed as
\begin{eqnarray}
& & \ket{\Psi_{N,M}^{\rm ini}} \nonumber \\
&=& \left(\bigotimes_{m=1}^{N+2M}\ket{\Phi^+}_{A_mB_m}\right)\otimes\ket{\xi}_{Y_1Y_2\cdots Y_{N+M}} \nonumber \\
&=& \left(\bigotimes_{m=N+1}^{N+2M}\ket{\Phi^+}_{A_mB_m}\right) \otimes \frac{1}{\sqrt{2^N}} \sum_{k_1,k_2,\cdots,k_{N}=0}^{1} \nonumber \\
& & \qquad y_{k_1,k_2,\cdots,k_{N}} \bigotimes_{i=1}^{N}\left(|00k_i\rangle+|11k_i\rangle\right)_{A_iB_iY_i} \otimes |\eta_{k_1,k_2,\cdots,k_{N}}\rangle_{Y_{N+1}\cdots Y_{N+M}}. \nonumber \\
\end{eqnarray}

After Bob's preparation, the state becomes
\begin{eqnarray}
|\Psi^{1}\rangle &=& {\mathcal{P}}_B(b_1,b_2,\cdots,b_N) \ket{\Psi_{N,M}^{\rm ini}} \nonumber \\
&=& \left(\bigotimes_{m=N+1}^{N+2M}\ket{\Phi^+}_{A_mB_m}\right) \otimes \frac{1}{\sqrt{2^N}} \sum_{k_1,k_2,\cdots,k_{N}=0}^{1} y_{k_1,k_2,\cdots,k_{N}}  \nonumber \\
& & \left\{ \bigotimes_{i=1}^{N} \left[(\ket{b_i}_{B_i}\bra{b_i})   C^{not}(Y_i,B_i)\right]
(|00k_i\rangle+|11k_i\rangle)_{A_iB_iY_i} \right\}
\otimes |\eta_{k_1,k_2,\cdots,k_{N}}\rangle_{Y_{N+1}\cdots Y_{N+M}}.
\end{eqnarray}
From \cite{wang06},
\begin{equation}
\left[(\ket{b_i}_{B_i}\bra{b_i})   C^{not}(Y_i,B_i)\right]
(|00k_i\rangle+|11k_i\rangle)_{A_iB_iY_i} 
= \sigma_{b_i}^{A_i} \ket{k_ib_ik_i}_{A_iB_iY_i}.
\end{equation}
So,
\begin{eqnarray}
|\Psi^{1}\rangle &=& \left(\bigotimes_{m=N+1}^{N+2M}\ket{\Phi^+}_{A_mB_m}\right) \otimes \frac{1}{\sqrt{2^N}} \sum_{k_1,k_2,\cdots,k_{N}=0}^{1} y_{k_1,k_2,\cdots,k_{N}} \nonumber \\
& & \left[\bigotimes_{i=1}^{N} \sigma_{b_i}^{A_i} \ket{k_ib_ik_i}_{A_iB_iY_i}\right] \otimes |\eta_{k_1,k_2,\cdots,k_{N}}\rangle_{Y_{N+1}\cdots Y_{N+M}}  \nonumber \\
&=& \left(\bigotimes_{m=N+1}^{N+2M}\ket{\Phi^+}_{A_mB_m}\right) 
\otimes \bigotimes_{n=1}^{N} \ket{b_n}_{B_n}
\otimes \frac{1}{\sqrt{2^N}} \sum_{k_1,k_2,\cdots,k_{N}=0}^{1} y_{k_1,k_2,\cdots,k_{N}} \nonumber \\
& & \left[\bigotimes_{i=1}^{N} \sigma_{b_i}^{A_i} \ket{k_i}_{A_i}\right] 
\otimes \left[\bigotimes_{j=1}^{N} \ket{k_j}_{Y_j}\right] 
\otimes |\eta_{k_1,k_2,\cdots,k_{N}}\rangle_{Y_{N+1}\cdots Y_{N+M}}.  \nonumber \\
\end{eqnarray}

After the teleportations from Bob to Alice, the state of qubits $Y_{N+1}\cdots Y_{N+M}$ are replaced by the qubits $A_{N+1}\cdots A_{N+M}$ \cite{six93}.
So, the state of qubits $A_1A_2\cdots A_{N+M}$ $Y_1Y_2 \cdots Y_{N}$ becomes
\begin{eqnarray}
|\Psi^{2}\rangle &=& \sum_{k_1,k_2,\cdots,k_{N}=0}^{1} y_{k_1,k_2,\cdots,k_{N}} \nonumber \\
& & \left[\bigotimes_{i=1}^{N} \sigma_{b_i}^{A_i} \ket{k_i}_{A_i}\right] 
\otimes \left[\bigotimes_{j=1}^{N} \ket{k_j}_{Y_j}\right] 
\otimes |\eta_{k_1,k_2,\cdots,k_{N}}\rangle_{A_{N+1}\cdots A_{N+M}}. \nonumber \\
\end{eqnarray}

After the step of Alice's sending, the state of qubits $A_1A_2\cdots A_{N+M}$ $Y_1Y_2 \cdots Y_{N}$ becomes
\begin{eqnarray}
|\Psi^{3}\rangle &=& {\mathcal{S}}_A(a_1,b_1,a_2,b_2,\cdots,a_N,b_N;x,G) |\Psi^{2}\rangle \nonumber \\
&=& \left(\bigotimes_{m=1}^N \ket{a_m}_{A_m}\bra{a_m}\right)
  \left(\bigotimes_{m=1}^N H^{A_m}\right)
  T^r_{N,M}(x,G)^{A_1A_2\cdots A_{N+M}}   \left(\bigotimes_{m=1}^N\sigma_{b_m}^{A_m}\right)
|\Psi^{2}\rangle \nonumber \\
&=& \left(\bigotimes_{m=1}^N \ket{a_m}_{A_m}\bra{a_m}\right)
  \left(\bigotimes_{m=1}^N H^{A_m}\right)
  T^r_{N,M}(x,G)^{A_1A_2\cdots A_{N+M}} \nonumber \\
& &   \sum_{k_1,k_2,\cdots,k_{N}=0}^{1} y_{k_1,k_2,\cdots,k_{N}} 
\left[\bigotimes_{i=1}^{N} \ket{k_i}_{A_i}\right] 
\otimes \left[\bigotimes_{j=1}^{N} \ket{k_j}_{Y_j}\right] 
\otimes |\eta_{k_1,k_2,\cdots,k_{N}}\rangle_{A_{N+1}\cdots A_{N+M}} \nonumber \\
&=& \left(\bigotimes_{m=1}^N \ket{a_m}_{A_m}\bra{a_m}\right)
  \left(\bigotimes_{m=1}^N H^{A_m}\right)
  T^r_{N,M}(x,G)^{A_1A_2\cdots A_{N+M}}  \nonumber \\
& &   \sum_{m=1}^{2^N} y_m
|m,D\rangle_{Y_1\cdots Y_N} \otimes |m,D\rangle_{A_1\cdots A_N} \otimes |\eta_{m}\rangle_{A_{N+1}\cdots A_{N+M}} \nonumber \\
&=& \sum_{m=1}^{2^N} y_{m} |m,D\rangle_{Y_1\cdots Y_N} \nonumber \\
& & \otimes \sum_{j=1}^{2^N} \left\{ \left[\bigotimes_{i=1}^N (|a_i\rangle_{A_i}\langle a_i|   H^{A_i} ) \right] \times |p_j(x),D\rangle \langle j,D| \right\} \times |m,D\rangle_{A_1\cdots A_N} \otimes G_j |\eta_{m}\rangle_{A_{N+1}\cdots A_{N+M}} \nonumber \\
&=& \sum_{m=1}^{2^N} y_{m} |m,D\rangle_{Y_1\cdots Y_N} 
\otimes \left[\bigotimes_{i=1}^N (|a_i\rangle_{A_i}\langle a_i|   H^{A_i} ) \right] \times |p_m(x),D\rangle_{A_1\cdots A_N} 
\otimes G_m |\eta_{m}\rangle_{A_{N+1}\cdots A_{N+M}}. \nonumber \\
\end{eqnarray}
Denote
\begin{equation}
|p_m(x),D\rangle_{A_1\cdots A_N}=\bigotimes_{i=1}^N |l_m^i(x)\rangle, \qquad (l_m^i(x)=0,1).
\end{equation}
Then,
\begin{eqnarray}
|\Psi^{3}\rangle &=& \sum_{m=1}^{2^N} y_{m} |m,D\rangle_{Y_1\cdots Y_N} 
\otimes G_m |\eta_{m}\rangle_{A_{N+1}\cdots A_{N+M}}
\otimes \left[\bigotimes_{i=1}^N \langle a_i| H |l_m^i(x)\rangle |a_i\rangle_{A_i} \right]  \nonumber \\
&=& \sum_{m=1}^{2^N} y_{m} |m,D\rangle_{Y_1\cdots Y_N} 
\otimes G_m |\eta_{m}\rangle_{A_{N+1}\cdots A_{N+M}}
\otimes \left[\bigotimes_{i=1}^N (-1)^{a_il_m^i(x)} |a_i\rangle_{A_i} \right]  \nonumber \\
&=& \left[\bigotimes_{i=1}^N |a_i\rangle_{A_i} \right] \otimes \sum_{m=1}^{2^N} 
\left[\prod_{k=1}^N (-1)^{a_kl_m^k(x)}\right] y_{m} |m,D\rangle_{Y_1\cdots Y_N} 
\otimes G_m |\eta_{m}\rangle_{A_{N+1}\cdots A_{N+M}}.
 \nonumber \\
\end{eqnarray}

After the teleportations from Alice to Bob, the state of qubits $Y_1\cdots Y_N$ $B_{N+M+1}\cdots B_{N+2M}$ becomes
\begin{equation}
|\Psi^{4}\rangle = \sum_{m=1}^{2^N} 
\left[\prod_{k=1}^N (-1)^{a_kl_m^k(x)}\right] y_{m} |m,D\rangle_{Y_1\cdots Y_N} 
\otimes G_m |\eta_{m}\rangle_{B_{N+M+1}\cdots B_{N+2M}}.
\end{equation}

Apparently,
\begin{equation}
{R}_N(x) \ket{m,D} = \ket{p_m(x),D}.
\end{equation}
So in the step of Bob's recovery, before the swapping oparations are implemented, the state of qubits $Y_1\cdots Y_N$ $B_{N+M+1}\cdots B_{N+2M}$ becomes 
\begin{eqnarray}
|\Psi^{5}\rangle &=& {\mathcal{R}}_B'(a_1,a_2\cdots,a_N;x) |\Psi^{4}\rangle \nonumber \\
&=& \sum_{m=1}^{2^N} \left[\prod_{k=1}^N (-1)^{a_kl_m^k(x)}\right] y_{m} 
\left(\bigotimes_{i=1}^Nr(a_i)^{Y_i}\right) |p_m(x),D\rangle_{Y_1\cdots Y_N} 
\otimes G_m |\eta_{m}\rangle_{B_{N+M+1}\cdots B_{N+2M}}  \nonumber \\
&=& \sum_{m=1}^{2^N} \left[\prod_{k=1}^N (-1)^{a_kl_m^k(x)}\right] y_{m} 
\left(\bigotimes_{i=1}^Nr(a_i)^{Y_i} |l_m^i(x)\rangle_{Y_i} \right)
\otimes G_m |\eta_{m}\rangle_{B_{N+M+1}\cdots B_{N+2M}}  \nonumber \\
&=& \sum_{m=1}^{2^N} \left[\prod_{k=1}^N (-1)^{a_kl_m^k(x)}\right] y_{m} 
\left(\bigotimes_{i=1}^N (-1)^{a_il_m^i(x)} |l_m^i(x)\rangle_{Y_i} \right)
\otimes G_m |\eta_{m}\rangle_{B_{N+M+1}\cdots B_{N+2M}}  \nonumber \\
&=& \sum_{m=1}^{2^N}  y_{m} \bigotimes_{i=1}^N |l_m^i(x)\rangle_{Y_i} 
\otimes G_m |\eta_{m}\rangle_{B_{N+M+1}\cdots B_{N+2M}} \nonumber \\
&=& \sum_{m=1}^{2^N} y_{m} |p_m(x),D\rangle_{Y_1\cdots Y_N}
\otimes G_m |\eta_{m}\rangle_{B_{N+M+1}\cdots B_{N+2M}}. \nonumber \\
\end{eqnarray}
After the swapping oparations, the final state of qubits $Y_1\cdots Y_{N+M}$ becomes
\begin{eqnarray}
\ket{\Psi_{N+M}^{\rm final}}_{Y_1Y_2\cdots Y_{N+M}} &=& \sum_{m=1}^{2^N} y_{m} |p_m(x),D\rangle_{Y_1\cdots Y_N} \otimes G_m |\eta_{m}\rangle_{Y_{N+1}\cdots Y_{N+M}} \nonumber \\
&=& T_{N,M}^{r}(x,G) \ket{\xi}_{Y_1Y_2\cdots Y_{N+M}}. \nonumber \\
\end{eqnarray}

Thus, we accomplish the proof.
\end{appendix}

\end{document}